# Calculations of Frequency Response Functions(FRF) Using Computer Smart Office Software and Nyquist Plot under Gyroscopic Effect Rotation

Hisham.A.Al-Khazali
PhD. Student in faculty of Science, Engineering and Computing,
School of Mechanical & Automotive Engineering, Kingston University, London, UK

Dr. Mohamad.R.Askari
Faculty of Science, Engineering and Computing,
School of Aerospace & Aircraft Engineering, Kingston University, London, UK

*Abstract* -Regenerated (FRF curves), synthesis of (FRF) curves there are two main requirement in the form of response model, The first being that of regenerating "Theoretical" curve for the frequency response function actually measured and analysis and the second being that of synthesising the other functions which were not measured,(FRF) that isolates the inherent dynamic properties of a mechanical structure. Experimental modal parameters (frequency, damping, and mode shape) are also obtained from a set of (FRF) measurements. The (FRF) describes the input-output relationship between two points on a structure as a function of frequency. Therefore, an (FRF) is actually defined between a single input DOF (point & direction), and a single output (DOF), although the FRF was previously defined as a ratio of the Fourier transforms of an output and input signal. In this paper we detection FRF curve using Nyquist plot under gyroscopic effect in revolving structure using computer smart office software.

## I. INTRODUCTION

Modal analysis has been used in the last two or three decades in many engineering discipline and technology fields to solve increasingly demanding structural dynamic problems,[1,2].Modal analysis has become a major technology in the quest for determining ,improving and optimization dynamic characteristics of engineering structure and has also discovered profound application for civil and building structure, biomechanical problems, space structures, acoustical instrument, transportation and nuclear plants, it can be measured by using :

*1-1 Computer simulation*

Such us computer–algebra simulation (CALS) is combination of symbolic and numeric methods, which is very well suited for efficient solving of complex problem,[3].The basis of (CALS) is mathematical models reproducing reality in sufficient detail, so (CALS) is independent of any specific field based on Mathematical. New and improved simulation techniques are in increasing demand; because development cycles need to be short end more and more. The computer–algebra simulation closes that gap by allowing a deeper insight into the physical functionality of technical process. With computer-algebra system numeric calculations of arbitrary precision can be done. This capability also covers all important mathematical field e.g. equations, integrals, differential equations, non linear fitting by using symbolic methods for numerical calculation. An example for this is the numerical solving of eq. (Newton's) method) or differential equations. (Calculation of Eigen values of the (Jacobi-matrix) for deciding if the equation is stiff or not),[4].The are many differences between (CALS) and other simulation techniques (e.g. numeric).

*1-1-1 Application of (CALS) method*

Is used in industry such us (air condition and heat technology process technology, vibration analysis, fluid mechanics, season technology. Control system, vacuum technology),[3,5].

*1-2 Computer smart office software*

The Smart Office is the software which is used in this project,[6]; The Smart Office Analyzer is suitable for accurate and efficient noise and vibration measurements, third-party data import/export, data analysis and reporting of the results. The SO Analyzer supports a wide range of measurement front–end (USB, PCI, PX1, and VX1) which enables the applications from (2 to 100) of input channels.
As the smart office software is used to carry out the frequency response functions (FRFs) and mode simulations of each setup, to measure the dynamic characteristic of a system or structure, a measure of the output response behaviour to measurable input excitation into the system has to be calculated, as it can be seen from the Figure (1), by dividing the output response signals by the input excitation signal, it is possible to determine the transfer function of the system.(Usually the transfer function measures in Laplace domain and frequency response in frequency domain).

Dr.Mohamad.R.Askari[2]
[2] Faculty of Science, Engineering and Computing,

The input signal changes as it travels through any dynamic system .Therefore ,the ratio of the output signal to the input does contain information on the inherent properties of the system .In modal testing the function measured is the frequency response function .On the other hand ,the same function ,in most electronic and control is known as transfer function.

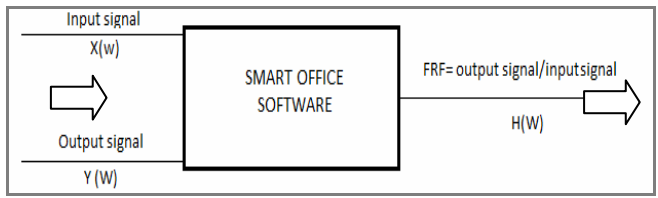

Figure 1. Signal processing.

*1-3 Measurement techniques*

 * Vibration testing (the maximum benefits are obtained when the instrumentation and analysis techniques used appropriate for and consistent with the desired test objectives for given machine and structure),[7], is often an important starting point in understanding that there may be problems with certain set of test data,[7,8] vibration tests are run for a number of reasons among them are:-

a- Engineering development testing.

b- Qualification testing.

c- Production screening testing.

d- Machinery condition monitoring.

* The same types of instruments, frequency analyzers and analysis methods are employed in nearly type of tests, [5]. Vibration type of vibration exciters, however, is employed for various tests.
In some cases the tests are run in the laboratory. In the other cases the only way to run a needed test is to run in it the field, under actual operating conditions , thus determining why a  particular  part of the structure is deteriorating too quickly, or why it is failing to function correctly under service conditions,[2,9],because of this wide range of technique and goals.
* There are many different type of vibration test. Some involved field measurements while the structure is in normal operational state, while other involves situations where the structure is excited by some external means, either in places in the field or in laboratory setting. These tests can be performed for a wide range of reasons such as vibration monitoring in order to determine a machine's suitability for operation, general vibration survey to find out what is happening, a complete motor analysis to determine the structures dynamics characteristics and so on,[10].

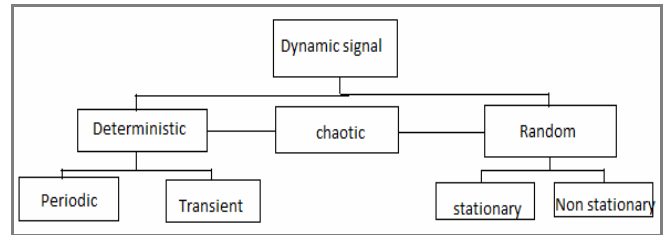

Figure 2. Dynamic signal classification,[11].

* In the study shown the natural frequency is depended on the initial orientation,[11].
* All linear system, the block diagram of general linear system show in Figure (3).

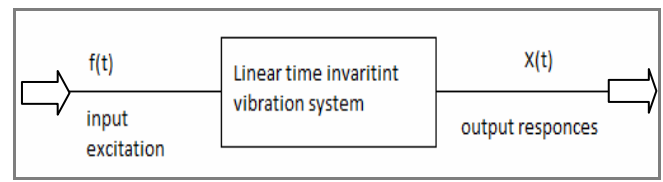

Figure 3. Input excitation and output responses.

*_ Transducer measurement considerations.
*_Vibration testing requires using transducers to measure motion as well as forces,[12,13].
*_Piezoelectric sending transducers are commonly used in accelerometer and forces gage,[14]. The popularity of piezoelectric sensors is done to their small size, high stiffness and high output. However piezoelectric sensors are charge generators that require special amplifiers called charge amplifier. This amplifier improper use can significantly affect our measurements,[14, 15].
*_The accelerometer response to sinusoidal and transient motion is explored,[2].

*1-3-1 The digital frequency analyzer*

  Real time analysis: real time analysis occurs when the data is proposed fast enough so that the data is sampled continuously at equal time increment,[13,15].

The effects of signal noise on (FRF) Measurements,[11].

$$\gamma^2 (\omega) = H_1 (\omega) / H_2 (\omega) = 1 / \{1+G_{nn} (\omega) / G_{ff} (\omega)\} \quad (1)$$

The coherence function is sensitive to the input signal's noise relative to the actual signal at each frequency ($\omega$). The decrease in coherence is dependent on the noise to signal ratio, [9,11].

$$IS/N(\omega)=\{G_{ff}(\omega)/G_{ff}(\omega)\}=\{ \gamma^2 (\omega) / (1- \gamma^2 (\omega)\} \quad (2)$$

• Good input signal to noise ratio results when the coherence is close to unity.

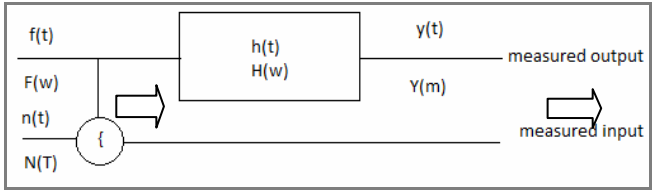

Figure 4. Linear model with noise (n(t)) in input signal.

The higher coherence, the closer together these two estimates are, however we must remember that while low coherences is an indication of measurement problems and high coherences is usually an indication of good quality measurements, there are situation where coherent noise in both measurements will give high coherences and poor measurements,[2,9]. Coherence is measured on a scale of(0.0 to 1.0),where 1.0 indicates perfect coherence is near the natural frequency of the system because the signals are large and are less influenced by the noise,[16]. Coherence values less than unity are caused by poor resolution, system nonlinearities, extraneous noise and uncorrelated input signals. Because coherence is normalized, it is independent of the shape of frequency response function, [13,17].

*1-3-2 Data analysis*

A subtle data processing problem that is associated with using digital techniques to simulate the derivative of single has been explored. The many of our data processing concepts come from the calculus of continues function but we calculate the results using discrete digital process,[2,8&18].

## II    NYQUIST PLOT

A Nyquist plot shows on the complex plane the real part of an FRF against its imaginary part with frequency as an implicit variable. The benefit of using Nyquist plot comes from the circularity of an FRF on the complex plane. This will be shown in Figure (5) for SDOF system with structural damping, we can draw Nyquist plot for its reacceptance, mobility and accelerance (FRF$_S$) Figure (6) three plots are not drawn to scale.

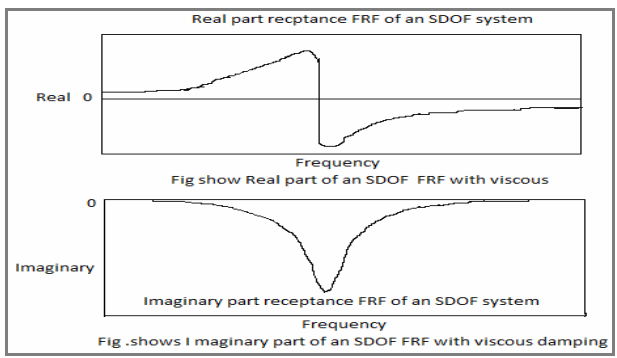

Figure 5. Real and Imaginary part of (SDOF) FRF with viscous damping,[2,19].

Although all three plots in Figure (6) appear to be circles, only the reacceptance FRF is real one. Form equations (1,2) we can see the reacceptance FRF begins from point $\{(k/(k^2+h^2)),((-h/(k^2+h^2)))\}$,while both mobility and accelerance FRF$_S$. Begin from origin. All three end at the origin. For measured FRF data only a finite frequency range is covered and a limited number of data point are available so that we always have a fraction of complete Nyquist plot,[2,18].

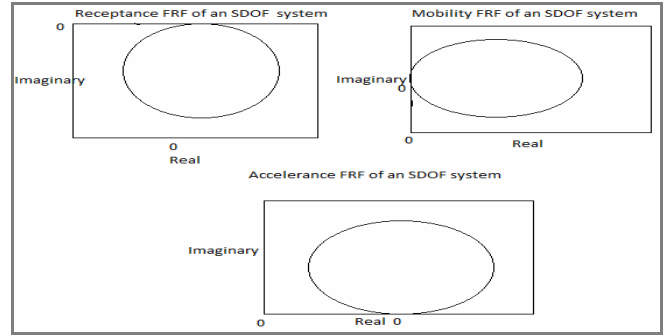

Figure 6. Nyquist plot, [8].

*2-1 Display and properties of an FRF of a damped MDOF system*

*2-1-1 Amplitude –phase plot and log-log plot*

The amplitude – phase plot of the FRF for a damped MDOF system consist of the plot of its magnitude versus frequency and that of its phase versus frequency,[18].

*2-1-2 Real and imaginary plots*

The real and imaginary plots consist of two parts: the real part of the (FRF) versus frequency and its imaginary part versus frequency,[2,8]. Real and imaginary plots are retracted to be its first part without damping. For MDOF system with structural damping, the real and imaginary parts can be derived analytically as,[5,8]:

$$\text{Re}(\alpha_{jk}(\omega)) = \text{Real}\left(\sum_{r=1}^{n} \{(\Phi_{jr}\ \Phi_{kr})/(\lambda_r - \omega^2)\}\right) \quad (3)$$

$$\text{Im}(\alpha_{jk}(\omega)) = \text{Imag}\left(\sum_{r=1}^{n} \{(\Phi_{jr}\ \Phi_{kr})/(\lambda_r - \omega^2)\}\right) \quad (4)$$

Figure (7) show the real and imaginary plots of an (FRF) for the 4DOF system,[2,19].

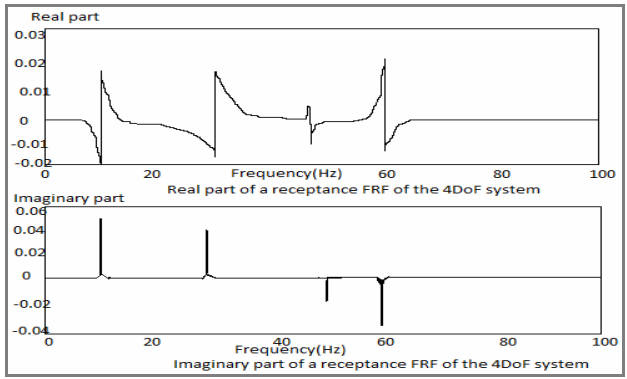

Figure 7. Real and Imaginary of FRF for (4DOF) system,[19].

*2-2 The relation between Nyquist plot and (FRF)*

The main benefit of using the Nyquist plot for an SDOF (FRF) comes from its circularity property in the complex plane. This still valid for a damped MDOF system.The circularity property does not exactly apply here since any vibration mode will be influenced by other modes of the system, thus compromising the simplicity form of an SDOF (FRF).Thus, the Nyquist plot is still one of the most useful plots for a damped MDOF (FRF). Figure(8)show the Nyquist plot of an (FRF)of the (4DOF) system,[8,19].The data point s do not connect full circles because of frequency resolution.

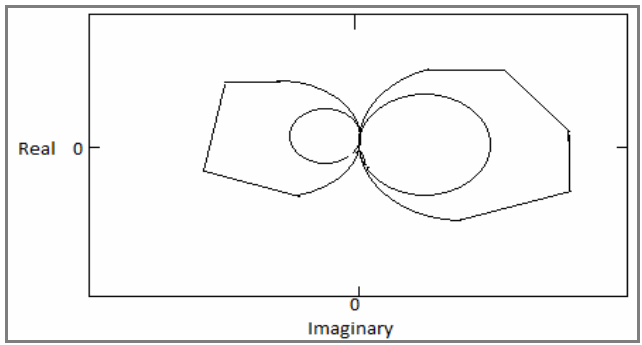

Figure 8. Nyquist plot of an FRF of the 4DOF system,[2].

*2-3 Modal testing*

Modal testing is in effect the process of constricting mathematical model to describe the vibration properties of a structure based on test data rather than conventional theoretical analysis,[8,14]. As such, it has application throughout the entire field of engineering. However; a relatively high level of theoretical competence is expected in order to understand properly the implications and limitations of the different types of measurement method-since, random and transient excitations and the like. We setup the gyroscopic modal testing show in picture (1).

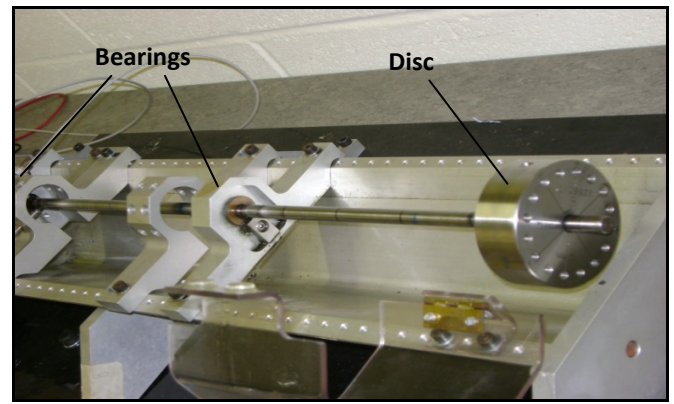

Picture (1) Experimental setup for the gyroscopic modal testing.

*2-3-1 Alternating acquisition and testing in experimental*

The Rotate acquisition and analysis software package is part of (m+p international's). It is designed for troubleshooting and analyzing noise or vibration problems related to the speed characteristics of rotating or reciprocating components of a machine in operation,[20,21].

*2-3-2 Test setup*

The rotor consisted of a shaft with a nominal diameter of 10 mm, with an overall length of 610 mm. The test rotor is shown in picture (1). Basically,. Included two journal bearings, RK4 rotor kit made by Bentley Nevada (the advanced power systems energy services company), could be used to extract the necessary information for diagnostic of rotating machinery, such as turbines and compressor. Been testing the process will be conducted on the rotary machine as the project is based on rotary dynamics reach practical results for the purpose of subsequently applied machinery rotary by using (Smart office program), and then do the experimental testing using the impact test, installed fix two accelerometer(model 333B32), sensitivity (97.2&98.6) mv/g in Y&Z direction and roving the hammer(model 4.799.375,S.N24492) on each point for the purpose of generating strength of the movement for the vibration body and the creation of vibration for that with creating a computer when taking readings in public that he was dimensions and introducing it with the data within the program(Smart office),[10,22]. Configuration for testing on the machines with rotary machine the creation of all necessary equipment for that purpose with the design geometry wizard is shown in Figure (9).

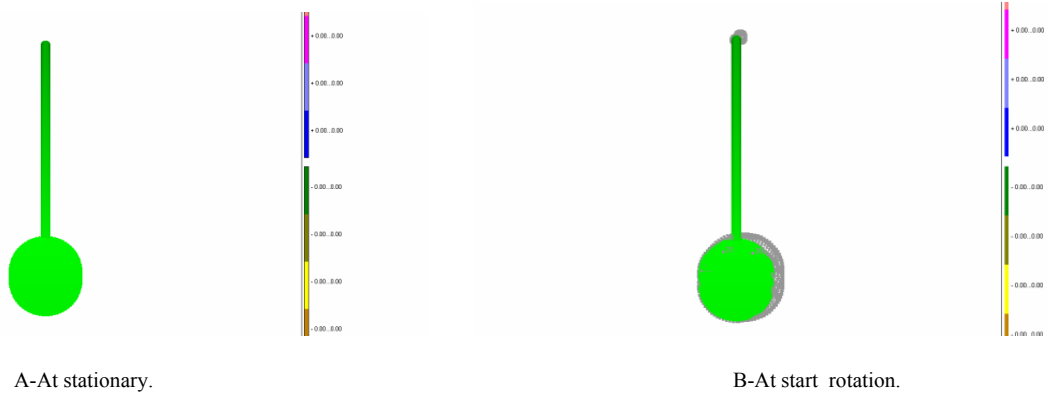

A-At stationary.    B-At start rotation.

Figure 9. Geometry design for modal (gyroscopic effect),experimental test using smart office;

*2-3-3 Whirls and stability*

As Eigen frequencies split with increasing spin velocity, ANSYS identifies forward (FW)&backward (BW) whirls, and unstable frequencies. Because of the orbit shape the shaft makes when rotating, this mode is sometimes referred to as a "Cylindrical" mode. If it is viewed from the front, the shaft appears to be bouncing up and down. Therefore this mode is also known as "bounce" or "translator" mode,[21]. Not to forget that mean while the rotor is also rotating. Therefore the whirling motion of the rotor (orbit shape path) can be in the same direction as the shafts rotation or can be in an opposite direction. This gives rise to the labels "forward whirl" FW and "backward whirl" BW. The rotor cross sections over the course of time for both synchronous forward and synchronous backward whirl.Not that for forward whirl, a point on the surface of the rotor moves in the same direction as the whirl.Thus, for synchronous forward whirl(i.e. unbalance excitation), a point at the outside of the rotor remains to the outside of the whirl orbit, [18,21].

With backward whirl, on the other hand, a point at the surface of the rotor moves in the opposite direction as the whirl to the inside of the whirl orbit during the whirl.

III    RESULTS OF EXPERIMENTAL METHODS FOR NYQUIST PLOT USING SOFTWARE

Beside the FRF, one other method of presenting the frequency response is to plot the real part versus the imaginary part. This is often called a Nyquist plot or a vector response plot. This display emphasizes the area of frequency response at response and traces out a circle. For proportionally damped system, the imaginary part is maximum at resonance and the real part is zero, this is shown in Figure (10).

Damping can also be estimated from the spacing of points along the Nyquist plot from the circle, [8,11].

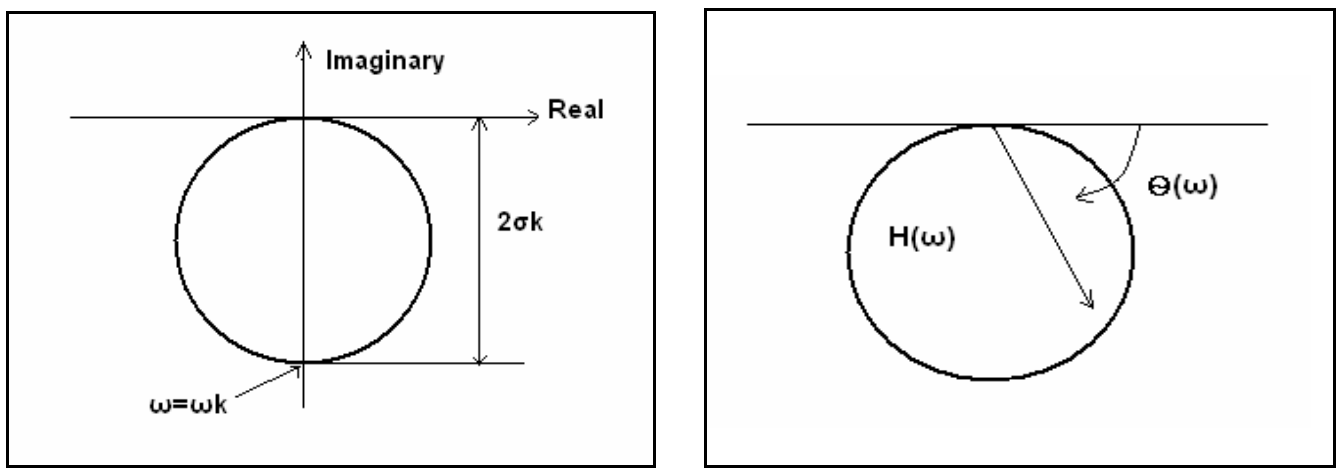

Figure 10. Nyquist plot of frequency response,[2,8&9].

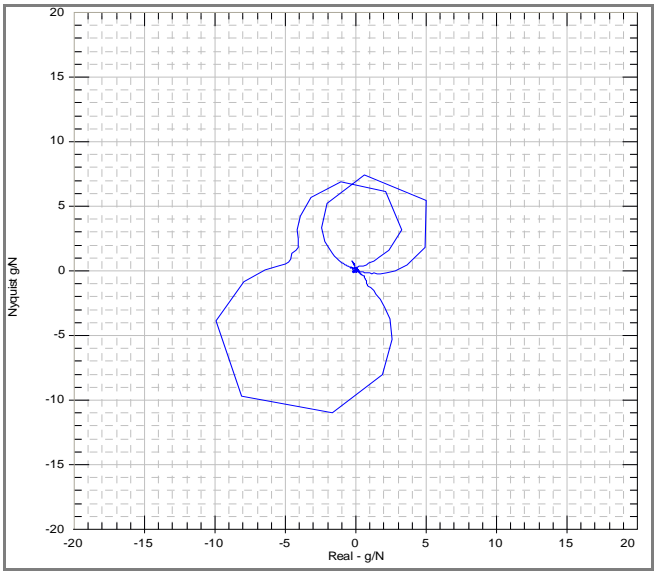

A-Speed of rotation (0)RPM,(stationary).

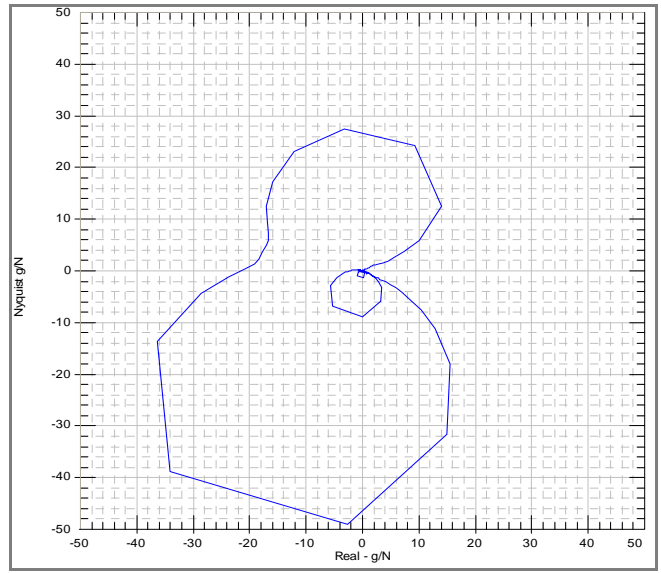

B- Speed of rotation (30)RPM.

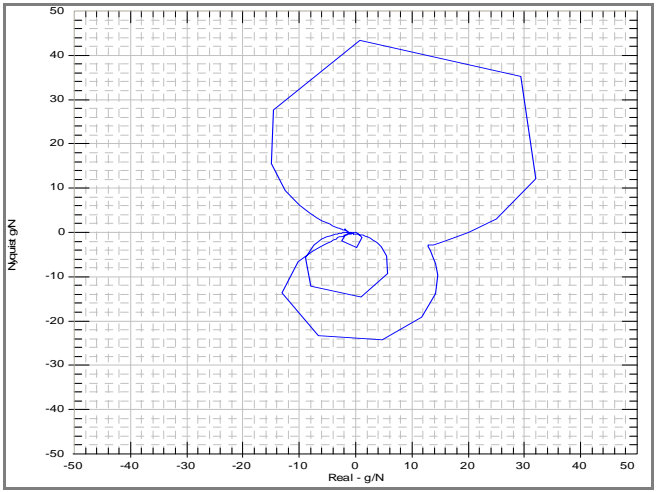

C-Speed of rotation (1000)RPM.

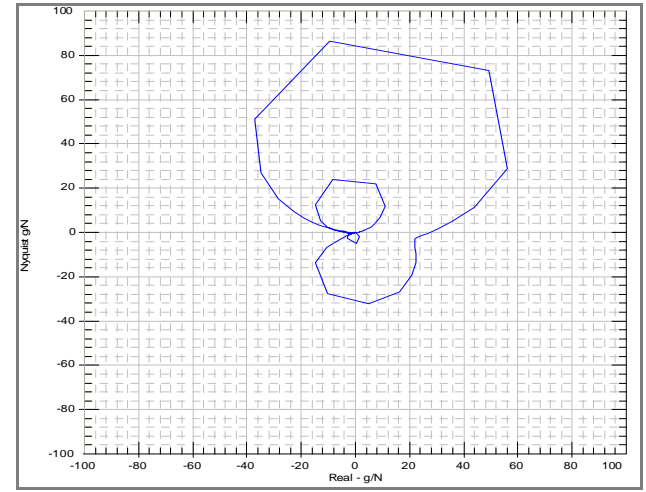

D-Speed of rotation (2000)RPM.

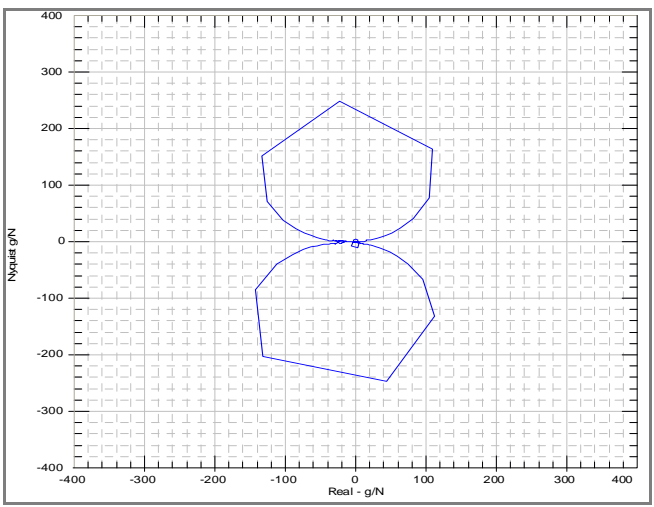

E-Speed of rotation (3000)RPM.

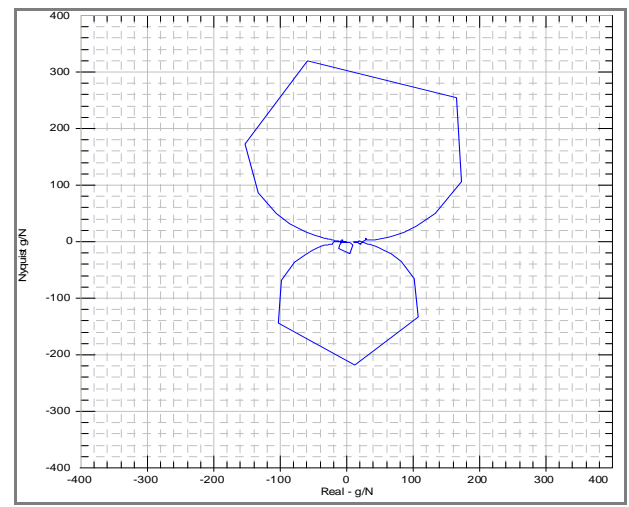

F-Speed of rotation (4000)RPM.

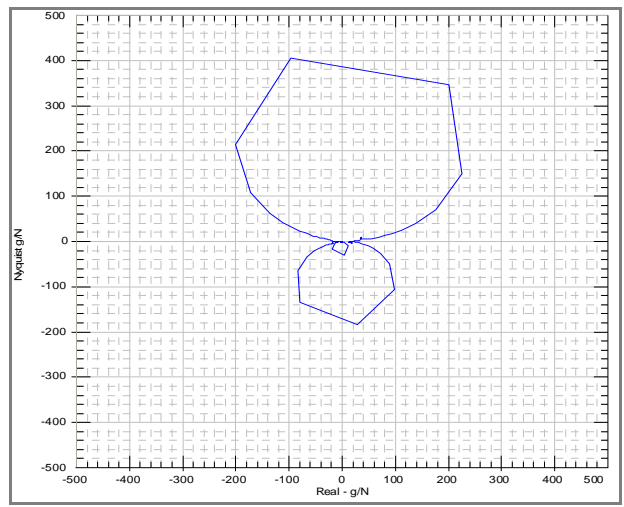

G-Speed of rotation (6000)RPM.

Figure 11. Experimental result show Nyquist plot (Frequency response function curve),Imaginary versus Real for different speed of rotation (Gyroscopic effect);

## IV DISCUSSION AND CONCLUSION

In Nyquist plot, as mentioned earlier, Figure (10), the imaginary part reaches a maximum at the resonant frequency and is 90° out of phase with respect to the input. Figure (11) show the experimental Nyguist plot of the system (Imaginary versus Real) for different speed of rotation at 0,30,1000,2000,3000,4000,and 6000 rpm respectively.
As it can be seen from the Figure (11,A,B,C,D) and as it was expected ,at 0,30,1000and 2000 rpm the Nyquist plot trace out a circle. This is a normal Nyquist plot display which is experimentally expected for any system. Having only one circular display in the Nyquist plot is due to the fact that there is only one natural frequency in the system at the selected rang of speed.

As mentioned before for this setup (Gyroscopic), as the speed increase, the natural frequency bifurcates into two; forward and backward whirl component frequencies. Because of this fact, the circular display of the Nyquist plot divides into two, so that the intersection of each circle to the imaginary axis present one natural frequency of the system.
As it is illustrated before, by increase of the rotor speed, the difference between forward and backward whirl frequency expands. As a result, at 6000 rpm, the Nyquist plot of the system shifts (turns) by 90° to put up with the increase of the natural frequencies, Figure (11,E,F,G) show this change.
The main benefit of using the Nyquist plot comes from its circularity property in the complex plane. The data point does not connect full circles because of frequency resolution we can see that clear in Figure (11,G).

## ACKNOWLEDGMENT

This study was supported by Iraqi Ministry of Higher Education and Scientific Research, Iraqi Cultural Attaché in London and Kingston University London. The authors would like to thank for supporting this work.

AUTHORS PROFILE

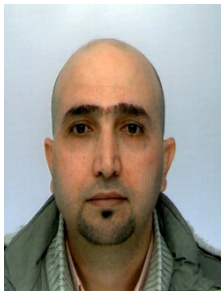

**Mr.Hisham.A.H.Al-Khazali**, He has PhD Student in Kingston University London. He was born in 28 Aug 1973 Baghdad/Iraq. Received his BEng in Mechanical Engineering (1996),University of Technology,Baghdad.MSc in Applied Mechanics, University of Technology, Baghdad (2000).

E-mail, k0903888@kingston.ac.uk

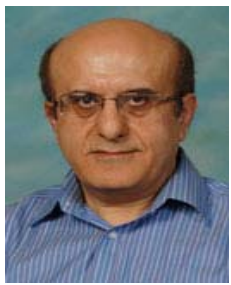

**Dr.Mohamad.R.Askari,** BSc(Eng), MSc, PhD, CEng, MIMechE, MRAeS. He has (Principal Lecturer, Blended Learning Coordinator),Member teaching staff in Kingston University London, His Teaching Area: Applied Mechanics, Aerospace Dynamics, Dynamics and Control, Structural and Flight Dynamics, Engineering Design, Software Engineering to BEng Mechanical and Aerospace second and final years.Year Tutor for BEng Mechanical Engineering Course and School Safety Advisor.

E-mail, M.Askari@Kingston.ac.uk